\documentclass[a4paper,10pt,oneside]{article}
\usepackage[english]{babel}
\usepackage{epsf}
\usepackage{amsmath}

\begin{document}

\title{ An optimal transient growth of small perturbations\\ in thin gaseous discs}

\author{ 
Razdoburdin D.N.$^{1}$,  Zhuravlev V.V.$^{2}$
} 

\date{ \it \small
1) Physics Department of the Moscow State University, Moscow, Russia 
\\
2) Sternberg Astronomical Institute, Moscow, Russia 
\footnote {E-mail for correspondence: v.jouravlev@gmail.com} \\
}

\maketitle


\bigskip

\begin{abstract}

A thin gaseous disc with an almost keplerian angular velocity profile, 
bounded by a free surface and rotating around point-mass gravitating object
is nearly spectrally stable. Despite that the substantial transient growth 
of linear perturbations measured by the evolution of their acoustic energy is possible. 
This fact is demonstrated for the simple model of a non-viscous polytropic thin disc of 
a finite radial size where the small adiabatic perturbations are considered 
as a linear combination of neutral modes with a corotational radius 
located beyond the outer boundary of the flow.
      
\end{abstract}

\vspace{1cm}


\newpage

\section{Introduction}

Accretion discs are observed astrophysical objects which 
can transform a high fraction of gravitational energy to thermal energy
and electromagnetic radiation.
The fact of effective conversion of the smooth large-scale rotation
to the small-scale dissipative chaotic structures indicates that 
the shear flows constructing an astrophysical discs must 
exhibit hydrodynamical activity, in other words, must be able
to generate complex motions that strongly increase the shear viscosity,
and cause the angular momentum transfer.   
Furthermore, the numerous observations of variabilities in discs  
that already endure the action of the accretion mechanism imply 
that these discs may also give birth to non-stationary large-scale 
deviations from the basic flow.

Theoretical analysis of the stability of astrophysical discs and dynamics 
of various perturbations in them has been performing through decades.
At present moment, however, the picture is not complete even for the simplest 
models of the shear flows. The same can be told about the linear 
theory when one considers a small perturbations only. In this case the obstacle 
is that the traditional eigenvalue approach to evolution of oscillations and
waves in shear flows when it is assumed that the solution has the form of definite frequency 
Fourier components $\propto e^{i\omega t}$ has revealed to be ineffective 
while studying the initial value problems. It turned out that even in the flows which 
are stable in a sense that all perturbation modes are exponentially damping
it is possible to set up such initial deviation from the background that 
it will be followed by a substantial transient growth of perturbations on the dynamical 
time-scale. Mathematically, this is closely related to the non-orthogonality of modes
in any physically meaningful metric space associated with them or, equivalently, to
the non-normality of the appropriate linear dynamical operator.

One of the first reviews on this subject has been done by Trefethen et al. (1993)
and the studies of certain models were carried out in the framework of
non-modal approach by Butler \& Farrell (1992), Farrell \& Ioannou (1993), 
Henningson \& Reddy (1994), Hanifi et al. (1996) and many others. The necessary 
references can be found in book by Schmid \& Henningson (2001) and in the latter 
review by Schmid (2007). 
Among other things, the so-called 'bypass mechanism' for the onset of turbulence
has been formulated. The concept states in general that the linear transient 
growth of optimal perturbations is a physical phenomenon responsible for 
the energy transfer between the large-scale motion and stochastic vortices.
Apart, the interaction between vortices themselves plays only a conservative 
role producing the stationary phase space distribution of vortices which 
is required to preserve modes taking part in energy transfer.

In astrophysical community the work by Ioannou \& Kakouris (2001) 
paid more attention to new ideas on 
the non-modal approach in hydrodynamical theory of perturbations. The authors 
analysed the spectrally stable two-dimensional flow of incompressible 
fluid rotating with the keplerian angular velocity profile and showed
that a stochastic external forcing in a form of white noise leads to 
the formation of strongly increased coherent structures that 
carry an angular momentum outwards.

It fact, the first studies in the area were published yet in the 1980s 
but have been performed in a slightly different form . We are talking about 
papers by Lominadze et al. (1988) and Fridman (1989), where the local 
evolution of an arbitrary initial vortical perturbations was considered. 
The local analysis of an astrophysical disc dynamics has been utilized since 
the work by Goldreigh \& Linden-Bell (1965).
In that framework assuming the constant shear in the radial direction it makes sense 
to change to co-moving shearing coordinate system to implement calculations for 
the spatial Fourier harmonics (SFH) of perturbations. Thus, one analytically 
obtains the evolution of all possible initial perturbations and as well as the 
behaviour of their part exhibiting transient growth. 
Let us remind that according to Fjortoft's theorem (1950) the constant shear 
flow constructed of incompressible and non-stratified fluid is asymptotically 
stable since it has no vorticity extremum. 

We would also like to mention Chagelishvili et al. (1996) who proposed a qualitative 
explanation of the transient growth phenomenon and Chagelishvili et al. (1997) 
who discussed a peculiar transformation of vortical SFH to the sound waves. 
In the latter and several subsequent papers (Tevzadze et al. 2003, Bodo et al. 2005, 
Tevzadze et al. 2008, Tevzadze et al. 2010) in addition to the energy extraction 
from the background a new process of the linear coupling of SFH is examined. 
The linear coupling of SFH leads to the mutual energy exchange between SFH of 
different types that correspond to fluid properties involved in calculations, namely, 
it's compressibility and stratification in the radial and vertical directions.
Here we can draw an analogy to the well-known resonant interaction 
(called 'coupling' as well) of modes in a critical layer where the pattern speed
of modes equals to the angular velocity of the flow (Cairns 1979). 

The local analysis in non-modal approach was also employed for different 
basic configurations by Yecko (2004), Afshordi et al. (2005), 
Mukhopadhyay (2005), Mukhopadhyay (2006), Johnson \& Gammie (2005a,b), recently by 
Volponi (2010) and others. 

Let us note, however, that despite the clear indications of the substantial transient 
growth found for a variety of disc models in the studies quoted above the transition to 
turbulence through bypass mechanism is seriously questioned 
by comprehensive numerical simulations performed 
by e.g. Lesur \& Longaretti (2005), Shen et al. (2006), Rincon et al. (2007), Lesur \&
Papaloizou (2010).
So the other ways to transition can't be excluded, for instance, by some kind of 
secondary instability (see the recent paper by Mukhopadhyay \& Kanak 2011).  

Nevertheless, the non-orthogonality of modes may become crucial in the
non-stationary appearance of turbulent viscous astrophysical discs. 
The studies by e.g. Umurhan et al. (2006), Regev (2008) and Rebusco et al. (2009) (see
also Shtemler et al. 2010) provide some evidence for that. They considered thin accretion
discs with an $\alpha$-parametrization  of viscosity which revealed a high 
transient growth of perturbations of density and vertical velocity component caused 
by the specially selected initial oscillations of the velocity in the disc plane. Note 
that in contrast to numerous investigations done in the framework of local analysis 
these studies involved large-scale vertical and radial structure of the disc.  

\vspace{0.5cm}

In this paper we concerned with the global optimal adiabatic perturbations, i.e. perturbations
exhibiting the highest possible transient growth in the fixed time interval. We 
would like to set up perturbations in thin non-viscous disc with a nearly
keplerian rotational profile. A simple baratropic configuration of a finite radial size
bounded by a free surface where the pressure vanishes is an object for our 
considerations. 
Due to small corrections set by the pressure gradient angular velocity, $\Omega$, 
will decrease slightly steeper than the keplerian one, $\Omega_K\propto r^{-3/2}$, 
as the radial distance, $r$, grows.
Lots of calculations by means of an eigenvalue analysis (see e.g. 
Kojima 1989 and Hawley 1990 with references therein) in such configuration
show that it has an exponentially growing modes, though the increments
become noticeable only for considerable pressure gradient in the flow when the 
rotation is essentially non-keplerian. 

Present study is motivated by the previous work by Zhuravlev \& Shakura (2009, 
ZS hereafter) where the optimal perturbations were found for a similar 
two-dimensional flow with a power law rotational profile $\Omega \propto r^{-q}$, where $3/2<q<2$.
It was revealed that, first, the strongest transient growth was produced by 
the linear combinations of slow sound modes with a corotation radius beyond 
the outer boundary of the flow, and second, the transient growth intensified 
while $q\to3/2$, i.e. while the rotation tends to keplerian one. 

Hence, we would like to examine the flows with an almost keplerian rotation what leads 
to a small $\delta = H/r$, where $H$ and $r$ are the characteristic half-thickness and radial
size. The smallness of $\delta\ll 1$ makes it possible to find the disc sound modes 
analytically employing the WKBJ scheme. As ZS we will extract optimal perturbations 
from span of the linear combinations of sound modes.

\section{Stationary flow}

\subsection{Basic assumptions}

Let us consider the axisymmetric flow in an external Newtonian gravitational potential 
$\Phi \propto -1/R$, where $R=(r^2+z^2)^{1/2}$ is distance from gravitating body  
and we will use the cylindrical coordinate system below
$\{r,\varphi,z\}$. 
We confine ourselves to the case of perfect fluid and polytropic configuration with 
an equation of state $p=K\rho^{1+1/n}$, where $p$ and $\rho$ are the pressure and 
mass density of material and $n$ is the polytropic index. 
Introducing enthalpy $h = \int dp/\rho$ we write equations of motion 
for the rotating stationary flow

\begin{equation}
\label{stat_system}
\begin{aligned}
&\frac{\partial h}{\partial r} = \Omega^2 r - \frac{\partial \Phi}{\partial r}\\
&\frac{\partial h}{\partial z} =  - \frac{\partial \Phi}{\partial z}, \\
\end{aligned}
\end{equation}
where in our model $\Omega = v_\varphi/r$
can be an arbitrary function of the radial coordinate
$r$. Following the previous studies we set it to be a power law 
\begin{equation}
\Omega = \Omega_0 \left ( \frac{r}{r_0} \right )^{-q}
\end{equation}
Here $r_0$ is the distance from the central body 
at which the rotation has a keplerian angular velocity
$\Omega_0$ in the midplane $z=0$.

Equalities
(\ref{stat_system}) allow us to find the distribution of enthalpy in the flow 

\begin{equation}
\label{h_profile}
h = (\Omega_0 r_0)^2  \left [ (x^2 + y^2)^{-1/2} - x_1 ^{-1} + 
\frac{1}{2(q-1)} \left ( x_1^{-2(q-1)} - x^{-2(q-1)} \right )  \right ],
\end{equation}
where the dimensionless coordinates are introduced
$x\equiv r/r_0$, $y\equiv z/r_0$. 
In this model of the basic configuration it's free boundary corresponds 
to the surface $h=0$ and $x_1<1$ is the crossing of that surface by an 
equatorial plane $y=0$. It is easy to show that in some range of the values of $q$ 
there is always a second crossing $x_2>1$. 
Let us call $x_1$ and $x_2$ simply the inner and the outer boundary of the flow 
and $x_d = x_2-x_1$ it's radial size. Here we are concerned with the configurations 
of a finite radial size $x_d<\infty$.

\subsection{Thin disc}

The situation when the rotational profile is almost keplerian is the most interesting to us.
Let $q = 3/2 + \epsilon^2/2$, where $\epsilon \ll 1$. Then, using the smallness of 
$\epsilon$,
and retaining in (\ref{h_profile}) only the main terms in orders of $y^2$ and $\epsilon^2$, 
we get an expression for the shape of the flow boundary, i.e. it's half-thickness profile
\begin{equation}
\label{H_x}
H(x) = 2^{1/2}\, \epsilon \, x\, [ \, 1 + \ln x - (x/x_1)(1+\ln x_1)\, ]^{1/2} 
\end{equation}

Looking at (\ref{H_x}) one can see that for moderate radial size 
$x_d\sim 1$ 
the half-thickness is controlled by the value of $\epsilon$
and we have stationary configuration in a form of the thin disc
with $H/x_d \ll 1$.

Entering a new variable $\delta \equiv H(x=1)$ which is approximately the maximum 
half-thickness in disc we finally have

\begin{equation}
\label{stat_struct}
\begin{aligned}
& H = \delta \, x \left [  \frac{ x_1 (1+\ln x) - x (1+\ln x_1) } {x_1 - 1 - \ln x_1 } \right ]^{1/2} \\
& h = \frac{H^2}{2 x^3} \left [ 1 - \left( \frac{y}{H} \right )^2 \right ], \\
\end{aligned}
\end{equation}
where enthalpy is dimensionless.

The relations
(\ref{stat_struct}) completely specify the polytropic disc where we are going to 
impose perturbations.

\section{Spectral problem}

The spectral problem implies calculation of definite frequency eigen-modes of perturbations 
for the specified stationary flow.
Non-viscous perturbation modes of small amplitude, 
$\propto exp(-i\omega t + m\varphi)$, 
which obey the equation of state common with the basic flow 
are described by the following equation

\begin{equation}
\label{equation}
\frac{D}{x\rho}\frac{\partial}{\partial x}\left(\frac{x\rho}{D}\frac{\partial W}{\partial x}\right)
-\frac{D}{\rho\,\bar\omega^2}\frac{\partial}{\partial y}\left(\rho\frac{\partial W}{\partial y}\right)
-\left[\frac{2m}{\bar \omega}\frac{D}{x\rho}\frac{\partial}{\partial x}\left(\frac{\Omega\rho}{D}\right)+\frac{D}{a^2}+\frac{m^2}{x^2}\right]W=0
\end{equation}
Here
$W=\bar p/\rho$ is the function to be determined, $\bar p(x,y)$ is 
a Fourier amplitude of the pressure perturbation, $a$ 
is a sound speed in the mean flow.
 
In (\ref{equation}) the following characteristic frequencies and its combinations have been 
assigned:
the shifted frequency of mode $\bar\omega=\omega-m\Omega$, 
the epicyclic frequency squared $\kappa^2=\frac{2\Omega}{x}\frac{d}{dx}\left(\Omega x^2\right)$, 
and the combination $D=\kappa^2-\bar\omega^2$.

Derivation of
(\ref{equation}) has been discussed by many authors, see e.g.
Goldreigh (1986), 
Kojima (1989), Kato (1987), one may also read a review by Kato (2001) on the oscillations in 
thin discs.

The boundary condition at the disc surface goes from equality to zero of the Lagrangian 
perturbation of pressure, which in turn is equivalent to regularity condition 
for $W$ at the boundary

\begin{equation}
\label{boundary}
W + \frac{1}{D} \frac{\partial h}{\partial x} \left ( \frac{2 m \Omega}{x \bar \omega}\, W - 
\frac{\partial W}{\partial x} \right ) 
+ \frac{1}{\bar \omega^2}\frac{\partial h}{\partial y}\frac{\partial W}{\partial y}=0
\end{equation}

A complete solution of the system
(\ref{equation}-\ref{boundary}) and the determination of the whole spectrum
of disc eigen-modes is a challenge for anyone. 
An eigen-frequency $\omega$ is in general a complex number what corresponds to exponentially 
growing and damping oscillations.

It is well known fact that either growth or attenuation of the particular modes 
take place only if the corotation point, $x_c$, where $Re(\omega)=m\Omega$, is located 
inside the flow since energy exchange either between the modes and the background 
or between the modes themselves having the total energy of opposite signs is possible solely
in the vicinity of $x_c$
(for details look in the book by Stepanyants \& Fabrikant 1996 and in the papers by Papaloizou \& 
Pringle 1987, Narayan et al. 1987, Glatzel 1987 with references therein)
The energy exchange in described by the resonant denominator in the coefficient of 
the basic equation
(\ref{equation}),
and the corotation point $\omega=m\Omega$ is it's singular point. 
The solution is multi-valued in the vicinity of that singular point what causes the problem of 
choice of a physically relevant branch what can be done according to the Lin's rule (1945).
Additionally, Lindblad resonances given by $D=0$ are the removable 
singularities of (\ref{equation}) but they are also of certain importance in 
perturbation dynamics.

However, we are interested in the non-modal dynamics here so let us choose the 
part of the spectrum that will be easy to obtain. Specifically, we consider 
the modes rotating slower than the flow itself, i.e. with a corotation radius 
$x_c>x_2$.
As was pointed out in ZS that is these modes that are able to exhibit 
the most considerable transient growth. 
Note that every such mode is neutral and have a constant total energy. 
We assume additionally that Lindblad resonances are out of the flow as well.
(the corresponding condition will be formulated below)
Hence, in (\ref{equation}) the term $\propto a^{-2}$ in coefficient in front of $W$  
is large everywhere in disc.
This implies that the wavelength 
of perturbations $\lambda$ is small and comparable to the disc thickness $\lambda \sim H \ll x_d$.
As it was shown by Nowak \& Wagoner (1992), see also Kato (2001), 
equation (\ref{equation}) is separable in the lowest WKBJ order 
into horizontal and vertical parts coupled by the slowly varying function $\Lambda(r)$.

However, here we would like to do further simplification 
and consider perturbations with no nodes in the vertical direction, 
so that $W$ is the function of $x$ only. Formally, this means that we require 
the maintenance of the vertical hydrostatic equilibrium in perturbed motion. 
This is of course a questionable requirement that artificially restricts our investigation
especially taking into account that strict vertical equilibrium in the perturbed flow
is possible for certain type of modes in case of isothermal vertical structure 
of the disc. So the general case $W(x,y)$ should be considered in the subsequent study.

Thus, all mentioned assumptions allow us to use the WKBJ approximation 
to determine the profiles of inertial-acoustic modes on interval $(x_1,x_2)$.

\subsection{Calculation of modes}

We first integrate the equation for small perturbations over the vertical direction obtaining

\begin{equation}
\label{eq_3D}
\frac{D}{x\Sigma}\frac{d}{dx}\left(\frac{x\Sigma}{D}\frac{dW}{dx}\right)
-\left[\frac{2m}{\bar \omega}\frac{D}{x\Sigma}\frac{d}{dx}\left(\frac{\Omega\Sigma}{D}\right)+(n+1/2)\frac{D}{h_*}+\frac{m^2}{x^2}\right]W=0
\end{equation}

where
\begin{equation}
\label{sur_a}
\Sigma\left(r\right)=\int\limits_{-H}^{H}\rho dz \propto H \left ( \frac{H^2}{x^3}\right )^n
\quad \mbox{and} \quad
h_*=\frac{H^2}{2x^3}
\end{equation}
are the surface mass density of the disc
and enthalpy in the midplane of the disc correspondingly.

Note that after a change $n\to n-1/2$ provided that $x_d<1$ equation (\ref{eq_3D})
coincides with the analogous equation in two-dimensional problem
which arises if we assume that perturbations are localized near the midplane 
of the disc, i.e. if in (\ref{equation}) all background quantities are equated 
to their equatorial values. This result was used to check the analytical 
solutions given below with the numerical calculations performed for two-dimensional 
problem in ZS (see also Zhuravlev \& Shakura 2007). 

Then, for part of the spectrum we have chosen the inner Lindblad resonance must 
be located outside of the flow, so that

\begin{equation}
\label{neutral}
\omega < (m-1)\Omega(x_2),
\end{equation}
what is followed by
$D<0$ and the WKBJ approximation gives an oscillating solutions. 
Let us note that in the case 
$m=1$ the inner Lindblad resonance is shifted to $x=0$ in the non-relativistic limit, 
consequently, disc settles into the evanescent zone for any arbitrary small frequency.
Thus, we only consider modes with $m>1$.

Then, it is convenient to write the WKBJ-solution in the form

\begin{equation}
\label{wkb}
W = C_0 S_1 \cos (S_0 + \varphi_0),
\end{equation}
where

$$
S_0 = \int\limits_{x_1}^x \left ( (n+1/2) \frac{-D}{h_*} - \frac{m^2}{x^2} \right )^{1/2} dx,
$$

$$
S_1 = \left ( \frac{-D}{x\Sigma} \right )^{1/2}  
\left ( (n+1/2)\frac{-D}{h_*} - \frac{m^2}{x^2} \right )^{-1/4},
$$
and $C_0$ and $\varphi_0$ are constants specified by the boundary conditions. 
Note that following WKBJ scheme we retain the terms in two main orders of $\delta$
proportional to $\propto\delta^{-2}$ and $\propto\delta^{-1}$.
For that reason the first term in coefficient in front of $W$ as it stands in 
equation (\ref{eq_3D}) is absent in \ref{wkb})
\footnote{
at the same time we retain the term 
$\propto m^2$ since we may consider modes with $m>>1$ as well}. 
For similar reasons combination $D$ entering (\ref{wkb}) is evaluated for 
a pure keplerian rotation.

WKBJ solution (\ref{wkb}) is irregular in the boundary points
$x_1$ and $x_2$. On the contrary, we mentioned above that the regularity of
$W$ at $x_{1,2}$ is equivalent to the boundary condition. 
Let us change to new independent variable 
$\tilde x\equiv |x-x_{1,2}|$
and rewrite equation 
(\ref{eq_3D}) in case of $\tilde x \ll 1$.
Practically, we assume that all quantities in 
(\ref{eq_3D}) having finite values at $x_{1,2}$ are equal to these values 
and approximate the disc half-thickness which vanishes at $x_{1,2}$ by the main 
term in it's expansion over $\tilde x$ getting $H=H_{1,2}\tilde x^{1/2}$ where 

$$
H_{1,2} = \delta x_{1,2} \left | {\frac{\ln{x_{1,2}}}{1+\ln{x_{1,2}}-x_{1,2}}} \right |^{1/2}
$$ 

In this way we get 
\begin{equation}
\label{eq_border}
\tilde x\, \frac{d^2 W}{d\tilde x} + (n+1/2)\,\frac{dW}{d\tilde x} + 
E_{1,2}W=0
\end{equation}
where
$$
E_{1,2} = \frac{(2n+1) (-D_{1,2})\, x_{1,2}^3}{H_{1,2}^2}, \quad D_{1,2} \,\, 
\mbox{are the values of } D \mbox{ in } \,\, x_{1,2}
$$

The solution of (\ref{eq_border}) which is  regular at $x_{1,2}$ is the following

\begin{equation}
\label{solve_border}
W=C_{1,2}\,\tilde x^{-(2n-1)/4} \, J_{n-1/2} ( \tilde z ),
\end{equation}
where $\tilde z = 2 E_{1,2}^{1/2}\,\tilde x^{1/2}$. 

Since the denominator of $\tilde z$ contains small $\delta$, 
at some distance from the boundary points $\tilde z>1$ while $\tilde x <1$. 
In this domain we take for $W$ an approximate expression using the asymptotic expansion 
of the Bessel function in 
(\ref{solve_border}) getting

\begin{equation}
\label{bessel_approx}
W \approx C_{1,2} \,\tilde x^{-n/2} ({4\pi^2E_{1,2}})^{-1/4} 
\cos{\left(2 E_{1,2}^{1/2}\, \tilde x^{1/2} - n \,\pi/2\right )}
\end{equation}

Matching (\ref{bessel_approx}) 
for the expansion of the WKBJ solution (\ref{wkb})
in the vicinity of $x_1$ and $x_2$ 
gives the zero phase $\varphi_0 = -n \pi/2$ and 
the dispersion relation

\begin{equation}
\label{dispersion}
\int\limits_{x_1}^{x_2} \left ( (2n+1) \frac{-D x^3}{H^2} - \frac{m^2}{x^2} \right )^{1/2} dx = \pi (n + k),
\end{equation}
where $k$ is an integer number.

Thus, the profiles of modes are given by the expressions
(\ref{wkb}) and (\ref{solve_border}) taking into account the following relations for constants
\begin{equation} 
\frac{C_0}{C_1} = \left ( \frac{H_{1}^{2n+1}}{2\pi x_{1}^{3n-1} (-D_{1})} \right )^{1/2}, \, \quad
\frac{C_2}{C_1}=\left(-1\right)^k \left [ \left( \frac{x_2}{x_1}\right)^{3n-1}
\frac{D_2}{D_1}\left(\frac{H_1}{H_2}\right)^{2n+1} \right ]^{1/2}
\end{equation}

\section{Optimal growth of perturbations}

In present study we utilize common procedure to calculate an optimal growth 
$G(t)$ for a linear combination of modes (for details see sect. 3.1 ZS and references therein). 
As in ZS we measure the growth of perturbations by their total acoustic energy

\begin{equation}
\label{akustic_en_3D_1}
E_a=\frac{1}{2}\int\Sigma\left(\delta v_r^2+\delta v_{\varphi}^2+(n+1/2)\frac{\delta h}{h_*}\right)xdxd\varphi,
\end{equation}
where $\delta v_r$, $\delta v_\varphi$ and $\delta h$
are the Euler perturbations of radial and azimuthal velocity components and 
enthalpy and the integration over the whole flow is implied.

The slight difference in comparison with ZS is that we integrate the energy density over $z$
what results in the emergence (see eq. (6) of ZS) 
of an additional factor ${n+1/2}$ in term
$\propto h_*^{-1}$ and the volume density is replaced by the surface density (\ref{sur_a}).

The metrics ${\bf M}$ in the linear span of the particular set of modes 
is calculated numerically taking the analytical expressions for the dispersion
relation and profiles of $W$ obtained above.

Then, matrix of propagator acting on the initial perturbations 
is diagonal in the basis constructed of modes
$\mbox{\boldmath$\Lambda$} = diag\{exp(-i\omega_1),exp(-i\omega_2),...,exp(-i\omega_N)\}$, 
where $\omega_j$ are the frequencies of those modes.

With the metrics determined we change to orthonormal basis where 
$\mbox{\boldmath$\Lambda$}$ transforms into non-diagonal and non-normal 
matrix. It's norm squared equals to it's 1st singular value which is $G(t)$.

An example of $G(t)$ at time intervals of order of the sonic time 
$t_s \sim (\delta\Omega_0)^{-1}$
and at the longer periods in comparison with the dynamical timescale
$t_d \sim \Omega_0^{-1}$ is displayed in Fig. 1. 

\begin{figure}[!h]
\epsfxsize=13cm \begin{center}\epsffile{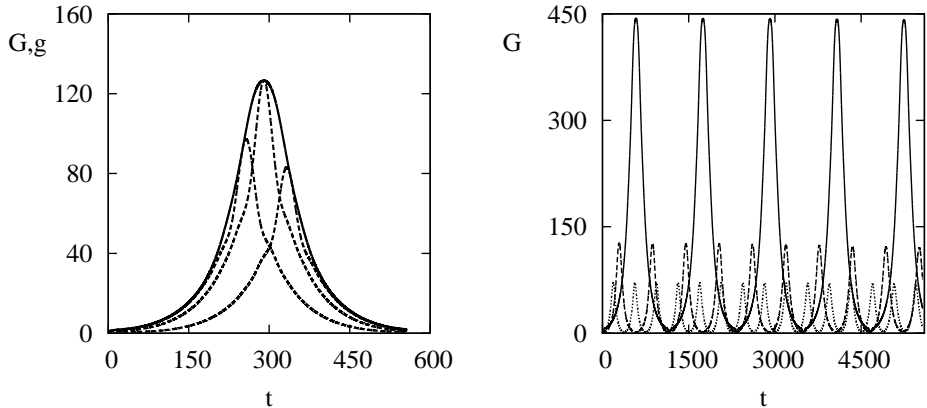}\end{center}
\caption{\small\rm 
Left-hand panel:
the solid curve represents an optimal growth $G(t)$ for a linear combination
of slow neutral modes of thin disc with $\delta=0.002$. The dashed curves 
correspond to the evolution of total acoustic energy $g(t)$ for certain combinations of modes.  
These combinations are optimal for the moments $t=250, 290, 390$ measured in 
the units of characteristic keplerian period $2\pi\Omega_0^{-1}$.
Right-hand panel: 
only the curves of $G(t)$ are plotted. The solid, dashed and dotted curves 
correspond to the values $\delta=0.001,0.002,0.003$.
On both panels: the dimension of a linear combination $N=20$, 
the other parameters are $x_d=1.0$, $m=25$, $n=3/2$. }
\end{figure}

Also one can see there the evolution of certain 
optimal combinations of slow modes, $g(t)$ (see eq. (10) of ZS).  
We draw time in units of the keplerian periods $t_0=2\pi\Omega_0^{-1}$ 
on the horizontal axis. Curves $g(t)$ are associated with perturbations 
that reach the highest possible (in linear problem) growth 
at $t=250,290,340$, when as it should be they touch with the overall curve 
corresponding to 
optimal growth $G(t)$.

As seen from Fig. 1 $G(t)$ has a quasi-periodical shape repeatedly reaching the peaks 
and perturbations increase by many times at intervals of several hundreds of $t_0$.
Clearly, this corresponds to sonic time-scale $\sim t_s$, 
and we keep in mind that even if there is a turbulent viscosity which 
switches on an accretion in disc the whole configuration 
evolves at the diffusion time-scale $t_\nu\sim \delta^{-2}t_d$. 
Thus, we conclude that the revealed transient dynamics in 
non-viscous rotating flow is possibly relevant to thin accretion 
discs at the intermediate time intervals in comparison with dynamical and 
viscous time-scales.

\subsection{Angular momentum flux}

Now we are concerned with the redistribution of the specific angular momentum 
during the transient dynamics. 
With the help of Savonije \& Heemskerk (1990) results let us write 
an appropriate expression for the angular momentum flux density of perturbations
in the radial direction (see sect. 4 of work by Savonije \& Heemskerk 1990 )

\begin{equation}
F = x\Sigma <\delta v_r \delta v_\varphi>,
\end{equation}
where brackets $<>$ mean an azimuthal average.

Note that $F$ is directly related to another important quantity describing 
the evolution of perturbations which is the flux density, $F_R$, 
of energy transferring from the background. 
$F_R$ is called also a Reynolds force and $F_R = -d\Omega/dx F$. 
Clearly, $F_R$ and $F$ have the same sign in our case, 
positive when angular momentum is transferred outwards and 
perturbations take energy from the background meanwhile (the epoch of $g(t)$ increase) and
negative in the opposite situation (the epoch of $g(t)$ decrease). 
Also note that following the spectral analysis it is easy to show that 
for any mode separately $F=0$ in the whole range between $x_1$ and $x_2$.

\begin{figure}[!h]
\epsfxsize=13cm \begin{center}\epsffile{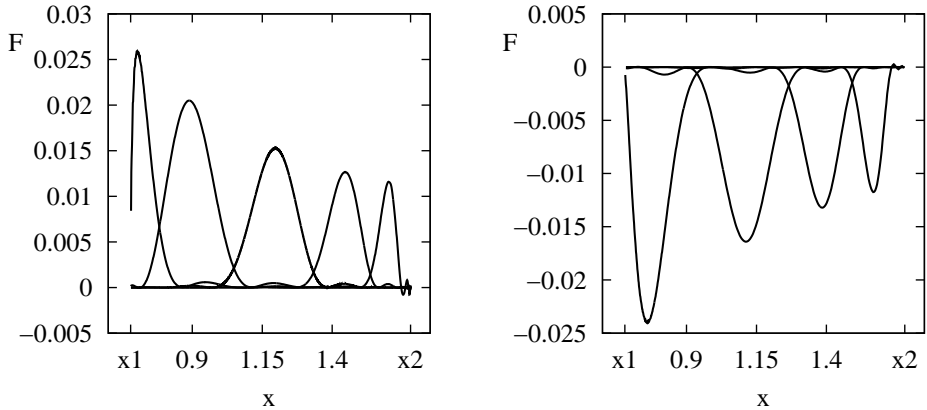}\end{center}
\caption{\small\rm
The radial shapes of azimuthally averaged angular momentum flux density $F$ of
transient perturbations are presented (Angular momentum density has only 
vertical non-zero component in our model). The curves correspond to evolution of a 
particular combination of modes which can be seen on the left-hand panel in Fig. 1 
as $g(t)$ for optimization moment $t=290$.
On the left-hand panel: the curves are obtained at $t=50, 100, 150, 200, 240$ before
$g(t)$ attains it's maximum value. Each profile has one high peak with it's position
shifting from the outer boundary to the inner boundary as time goes on. 
On the right-hand panel: the curves are obtained at $t=290, 350, 400, 450$ after
$g(t)$ attains it's maximum value. Each profile has, analogously, one deep 
minimum with it's position shifting from the inner boundary to the 
outer boundary as time goes farther.
On both panels: the dimension of a linear combination $N=20$, 
the other parameters are $\delta=0.002$, $x_d=1.0$, $m=25$, $n=3/2$. 
}
\end{figure}

It is quite straightforward to calculate $F$ profile for optimal perturbations. 
In Fig. 2 we see how the shape of $F$ transforms during the evolution of 
certain optimal combination of modes associated with the moment $t=290$ (when it 
touches the curve $G(t)$) in Fig. 1. 
The moment $t \simeq 290$ coincides with the local peak of
$g(t)$ in this particular case.  

It turns out that the specific angular momentum alters at rate varying along the interval $(x_1,x_2)$.
More definitely, the highest angular momentum flux and the most rapid alternation of 
angular momentum density is always localized in some narrow part of $(x_1,x_2)$.

As $E_a$ increases the mentioned domain of an intensive angular momentum outflow
shifts from the inner to the outer boundary of disc. 
After the transient growth attains it's peak the abrupt turnover 
happens with $F$ so that angular momentum starts to flow towards the inner 
boundary. The latter result is expected since we are dealing with 
the nearly stable configuration.
Clearly, the domain of the most intensive inflow of angular momentum 
appears first near the inner boundary and then moves to the outer boundary.

\subsection{Parametric study}

In this section we briefly discuss the dependence of $G$ on the free 
parameters of the problem. 
They are the azimuthal wavenumber $m$, the polytropic index $n$, the typical 
half-thickness of disc given by $\delta$ and the radial size of the disc $x_d$.
Besides, we have to fix the number of modes, $N$, involved in calculations, i.e. 
the dimension of the linear span for perturbations we consider. 
However, the numerical tests show that a kind of saturation of the 
transient growth occurs as $N$ increases. In other words, 
as we include more and more modes
the magnitude of $G$ stops to rise up noticeably 
from some value of $N$. This is illustrated by Fig. 3 where we present 
the dependence of magnitude of the first peak at $G$ curve, $G_{max}$, 
on $n$ obtained for different $N$. 
Note that this saturation upon reaching $N\sim 30-40$ takes place 
for the model of ZS as well. The polytropic index varies in the
range from $3/2$ to $3$ what means passing from non-relativistic to relativistic gas 
in case of the flow with a zero gradient of entropy. 
We conclude that $G_{max}$ is almost independent of $n$ 
\footnote{
Small variations of $G_{max}$ in Fig. 3 are actually caused by the fact 
that while moving through the parameter space one has to adjust slightly the 
set of neutral modes to satisfy the condition 
(\ref{neutral}). 
As should be noted we always take the first neutral modes as 
measured by the distance of their corotation point from $x_2$.  
}.

\begin{figure}[!h]
\epsfxsize=10cm \begin{center}\epsffile{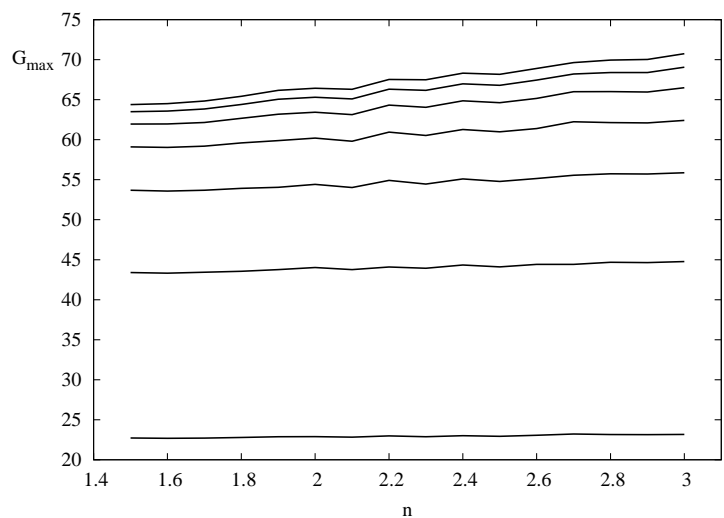}\end{center}
\caption{\small\rm
The first peak of an optimal growth curve, $G_{max}$, depending on the polytropic index,
$n$, calculated for various dimensions of a linear combination $N$. From 
bottom to top: $N=4,8,12,16,20,24,28$. The other parameters are $\delta=0.01$, 
$x_d=2.0$, $m=15$. 
}
\end{figure}

\begin{figure}[!h]
\epsfxsize=13cm \begin{center}\epsffile{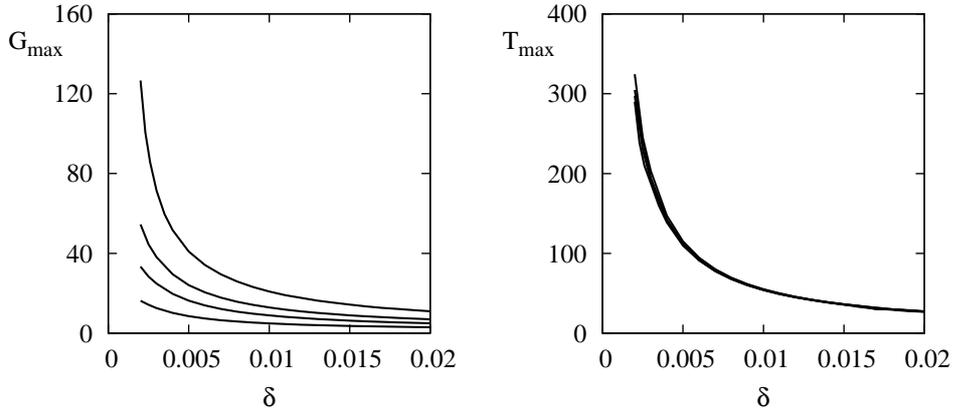}\end{center}
\caption{\small\rm
The dependence of $G_{max}$ (left-hand panel) and $T_{max}$ (right-hand panel) on 
$\delta$. 
On the left-hand panel: from low to upper curves $m$ has the values correspondingly
$m=5,10,15,25$. 
On the right-hand panel: the corresponding curves are superimposed on one another since 
the dependence of $T_{max}$ on $m$ reveals to be weak. 
On both panels: the other parameters are $N=20$, $x_d=1.0$, $n=3/2$. 
}
\end{figure}

\begin{figure}[!h]
\epsfxsize=13cm \begin{center}\epsffile{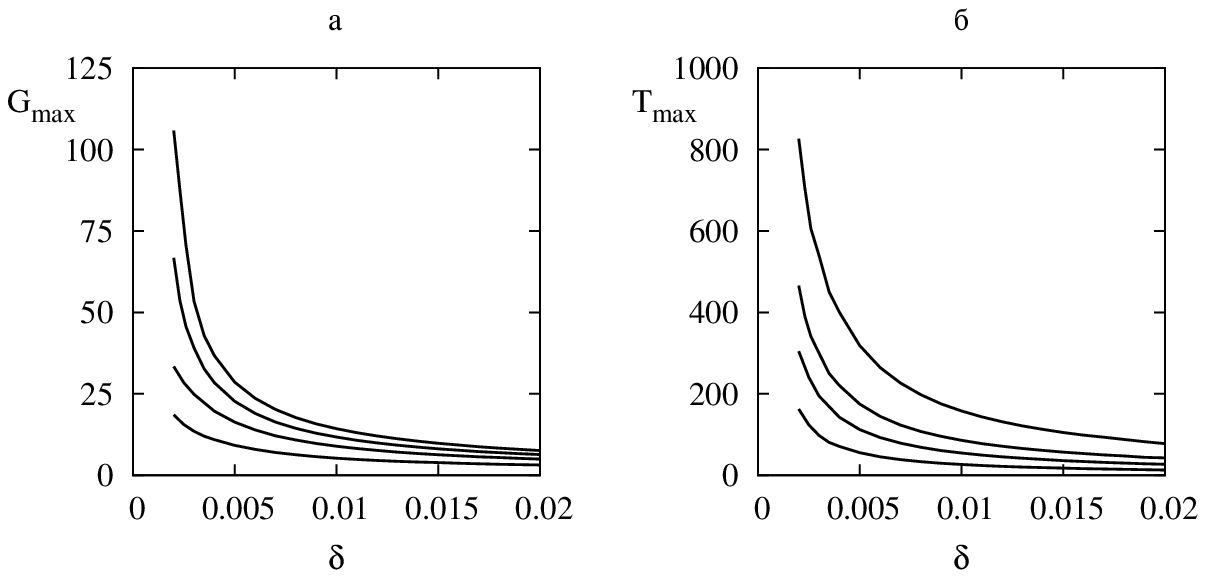}\end{center}
\caption{\small\rm
The dependence of $G_{max}$ (left-hand panel) and $T_{max}$ (right-hand panel) on 
$\delta$. 
On both panels: from low to upper curves $x_d$ has the values correspondingly 
$x_d=0.5,1.0, 1.5, 2.5$. The other parameters are $N=20$, $m=10$, $n=3/2$. 
}
\end{figure}

Then, in Fig. 4 we demonstrate how $G_{max}$ depends on $\delta$ which is the key 
parameter of our model. This is done for several values of $m$. 
On the right-hand panel of Fig. 4 one can see also the curve of $T_{max}(\delta)$ 
where $T_{max}$ is the time interval given by the condition $G(T_{max}) = G_{max}$. 
As we see $G_{max}$ increases inversely to the half-thickness of the disc
what is consistent with the results of ZS where the similar dependence on 
the Mach number was presented. Additionally, the substantial dependence 
on $m$ is confirmed here. Precisely, $G_{max}$ is proportional to the azimuthal 
wavenumber. Apart, the approximate duration of the transient growth given 
by $T_{max}$ alters weakly with $m$. 
At the same time concerning it's dependence on the half-thickness 
one can find the relation $T_{max}\sim (\delta\Omega_0)^{-1}$
corresponding to sonic time-scale what has been 
already discussed above.   

Finally, in Fig. 5 it is displayed how $G_{max}$ and $T_{max}$ alter with variation 
of $\delta$ for few values of $x_d$. As we see both quantities grow as $x_d$ increases
in agreement with ZS.

\section{Discussion}

In this paper we reveal the significance of the linear transient dynamics 
concept in application to astrophysical discs. We point out that in contrast to the
majority of studies in the area the global problem was considered, i.e. 
the perturbation dynamics was examined in the whole disc taking into account it's 
free boundary. Moreover, we employ the approximation of thin disc with 
an almost keplerian rotation. The latter is quite important in astrophysics.
However, several simplifications have been done to obtain the results. 
First, we neglect the viscosity in disc and assume it to be baratropic.
Then, we consider only perturbations that preserve the vertical 
hydrostatic equilibrium. And after all we make one more assumption 
taking the linear combinations of neutral modes only. We call these 
neutral modes ``slow modes'' since their corotation radii are beyond 
the outer boundary of the disc. 

Nevertheless, even the oversimplified model shows that thin disc is able to 
generate perturbations with a substantial transient growth. 
Further, to the contrary with the conclusions of modal analysis 
an optimal growth gets higher with the decrease of disc thickness $\sim\delta$, 
see Fig. 4 and Fig. 5. 
Let us remind that sonic instability, i.e. an exponential growth of 
sound modes generated by a similar toroidal configurations 
ceases promptly as one passes towards the keplerian rotation. 
Note again that the total energy of each neutral mode 
remains constant in time and azimuthally averaged radial component of 
angular momentum flux density corresponding to that mode equals to zero 
at any point of $(x_1,x_2)$. But nothing like that belongs to the linear 
combination of those modes which exhibits a considerable growth 
of an acoustic energy (see Fig. 1) accompanied by the increasing outflow of the specific angular 
momentum to disc periphery (see Fig. 2). 

We find that the transient growth acts on the time-scale $\sim(\delta\Omega_0)^{-1}$,
intermediate between the dynamical $\sim \Omega_0^{-1}$ and 
viscous $\sim\delta^{-2}\Omega_0^{-1}$ timescales. 
This implies that optimal perturbations can grow up much faster than 
disc structure changes under the action of viscous forces causing 
a gradual diffusive angular momentum transfer. 
Consequently, the transient dynamics phenomena can be important in accretion
discs that already have an enhanced turbulent viscosity. 
Though, one must keep in mind that in highly viscous discs the viscous 
forces should be allowed for changing the dynamics of perturbations themselves
since their characteristic scale $\lambda$ is small $\lambda \sim \delta$ and 
the corresponding time-scale is comparable to the dynamical one. 

The latter suggestion can be treated as one of the ways to develop the present 
study. Besides that, it would be possible to 
formulate problem on the stochastic forcing of the disc 
what was considered by Ioannou \& Kakouris (2001) for their shearing flow model. 
Note that we take into account compressibility of medium and the presence
of the free boundaries. 
However, perhaps the more important development of the research is to 
include the vertical motions in the perturbed flow, i.e. to include 
the dependence of perturbations on the vertical coordinate. The results 
of Rebusco et al. (2009) indicate that this can reveal even more 
intensive transient dynamics.

To conclude, let us remind that we take merely the modes without the turning 
points (Lindblad resonances) and the evanescent WKBJ-zone inside $(x_1,x_2)$. 
Additionally, the nearest Lindblad resonance is restricted to lie outside of the flow at some 
small distance $d_L\sim\delta$ beyond the outer boundary in order for WKBJ scheme 
to be valid everywhere in $(x_1,x_2)$ and to match smoothly a WKBJ-solution for an asymptotic 
expansion of a boundary solution close to $x_2$, see above.
However, the numerical tests showed that magnitude of the transient growth is 
sensitive to the presence of modes that have the mentioned Lindblad resonance the most close
to $x_2$. That is why it would be desirable to include into consideration 
modes with resonances inside the flow.

\vspace{0.5cm}

\noindent

This work was in part supported by ``Research and Research/Teaching staff of Innovative Russia''
for years 2009-2013 (State Contract No. P2552 on 2009 November 23)
and in part by a grant from the President of the Russian Federation for
the State Support of Young Russian PhDs (MK-73.2011.2)

\end{document}